# Security and Safety Aspects of AI in Industry Applications

Hans Dermot Doran
*Institute of Embedded Systems*
*ZHAW School of Engineering*
Winterthur, Switzerland
hans.doran@zhaw.ch

*Abstract*— In this relatively informal discussion-paper we summarise issues in the domains of safety and security in machine learning that will affect industry sectors in the next five to ten years. Various products using neural network classification, most often in vision related applications but also in predictive maintenance, have been researched and applied in real-world applications in recent years. Nevertheless, reports of underlying problems in both safety and security related domains, for instance adversarial attacks have unsettled early adopters and are threatening to hinder wider scale adoption of this technology. The problem for real-world applicability lies in being able to assess the risk of applying these technologies. In this discussion-paper we describe the process of arriving at a machine-learnt neural network classifier pointing out safety and security vulnerabilities in that workflow, citing relevant research where appropriate.

## I. Introduction

### A. Motivation

In industry circles Artificial Intelligence (AI), Machine Learning (ML) and Neural Networks (NN) have become established buzzwords and, as with many buzzwords, industry representatives are obliged to determine how the technologies can help or hinder their businesses and risk assessments on the introduction of these technologies to their companies and products. Many experienced industry representatives will have seen a number of buzzwords come and go and will be wary of ones that seem to attract a number of negative headlines. A well-known example of a negative headline is automated driving [1] where well-publicised failure modes [2] raise serious doubts over the technology domain.

Available answers to these questions are not confidence inspiring. Questions pertaining to fundamental understanding, reliability and explainability have been asked in the 1990's [3]. One 2014 publication on the observation that perturbance of input/training data can simply and drastically affect the outcome of a classification action [4] seems to have woken the broader research community up to the fact that professionalisation of the theme is required. Since then, explainable AI (XAI) and understandable AI have become unfortunately chosen buzzwords – an experienced industry representative might well ask how a theme can be researched and so publicly discussed since the early 80's and still not be understood.

False positive rates commonly and stubbornly occupy the 0.5-5% range so there can be no assumption of classifier-reliability. Such performance conditions effectively relegate the use of such classifiers to consultative and not determinative functions. In monetary terms, this means accepting the occurrence of false negatives (good product identified as faulty) to avoid false positives (faulty produce identified as good), thus increasing the costs of true positives.

These observations motivate this paper. We target the reader in the industrial domain seeking orientation points for further investigation in the context of her decision-making with regards to technology adoption.

### B. Methodology

Whilst largely a discussion paper, our interest is primarily in the domain of high integrity systems. We title the paper using the words safety and security. The underlying concepts are well-understood and well-defined, notably by specifications such as the IEC 61508 and IEC 62443. In implementation they both require the property termed "high-integrity" as a basis and given the similarities and respecting the differences between the two domains we see it as more constructive to frame such discussions as such, highlighting any disunions as necessary.

### C. Structure

This discussion-paper is structured accordingly, we finish this section by mentioning useful previous work. In Section II we then build a simple orientation-model of the classifier workflow. Using this workflow, we categorise activities and examine their vulnerabilities. In Section III we summarise and suggest future work.

## II. The AI / Machine Learning / Deep Neural Network Classifier Workflow

A neural network is an architecture which is one of a set of tools in machine learning which in turn is a subsection of artificial intelligence. To a certain extent, (deep) neural networks can be viewed as an attempt to overcome the limitations of alternative machine learning techniques, say statistical methods, in an attempt to achieve intelligent, informed might be a better word, behaviour in synthesised machines. By introducing a series of hidden (deep) layers, it is hoped that increasing abstractions of the input dataset can be drawn. By passing data thought to be representative of the problem domain through an algorithm (training), we arrive at a model (trained neural network) that can be executed and with which we hope to be able to classify previously unseen data. In this paper we are concerned with the class of deep neural network algorithms as classifiers.

*A. Neural Network Basics*

The architecture of neural networks is well known [5] and can be characterised as an attempt to replicate the functional activity of an organic brain. A neural network consists largely of artificial neurons which can be triggered depending on input triggers. The artificial neuron is generally represented as a multiply-accumulate (MAC) unit where a number of input weights are multiplied and accumulated to provide an output result which is later rectified. These MAC units are arranged in layers, the number of input MACs often corresponding to the number of sample points of the input – for instance pixels in an input image. The classifier output is separated from the input by a number of layers and an algorithm specific interconnection between the MACs of one layer to the MACs of the following layer. Intermediate layers can feature widths that are wider or narrower (dimension increase or reduction) than the previous layer. The output layer will generally be a width that corresponds to the number of discrete classifications and will generate for each, a number that represents a probability (or confidence) of the input being in this category. This can be further refined by an additional non-MAC layer (e.g. soft-max) that determines an absolute output of the classifier.

The weights for a particular problem, or data, set are established by a method known as back-propagation where the weights of the MAC inputs for a defined classification to a known input can be calculated. The use of labelled data (for instance images of cats, dogs, jars labelled as such) is known as supervised learning. Training time can vary, depending on the number and sizes of layers, the training sample set and of course the platform, from hours to weeks: the expectation is that graphics processing units (GPUs) are used to train. A percentage, 30% or so, of the possible input samples are reserved for the validation run and are passed to the trained model allowing the classification results to be evaluated. Should unacceptable classification results be achieved the training run can be repeated or the number of layers and/or their connectivity (algorithmic manipulations) adjusted so that more acceptable results are achieved. The convergence of this iterative process is thus largely dependent on the intuition and the experience of the implementer rather than any definable mathematical model. The reduction of this latter influence is the primary motivation behind automatedAI.

*B. Implementation Aspects*

In terms of high integrity, the process described above exhibits a number of serious intrinsic vulnerabilities. To explore these, we construct a simple model (see Fig. 1). We divided the process into three environments, namely the data set environment, the modelling environment and the target environment, and discuss the vulnerabilities of these environments in terms of integrity.

*1) Dataset Environment:* Datasets are, obviously, crucial to training models. There are a number of publicly available datasets in the domain of image recognition, available from various institutions and appear to be, disturbingly often, used uncritically by researchers. Well-known failures of facial recognition include minorities being misclassified due to a preponderance of white people in the dataset [6]. This misclassification is directly attributed to the actions of bias in choosing the images and their labels/classifications. The well-known ImageNet dataset [7] features, apparently, questionable and scientifically unacceptable categorisations such as "Slut", "Closet Queen" and the like [8][1]. An interesting discourse on the creation and curation of the ImageNet data base, elements of which are undoubtably representative of many of the publicly available datasets is offered by Denton et. al. [11] and a structured analysis of the differing types of bias is offered by [12]. Methods of harnessing ML to reduce the effects of bias have also been explored [13]. In typical industrial embedded applications, the use of publicly available data sets is expected to be limited and there appears to be an implicit assumption that in many applications it will be necessary to generate the dataset artificially. Nevertheless, awareness of the data-bias issue in particular and data curation in general is paramount to securing the integrity of the resulting model.

The integrity of the model can also be compromised by manipulation of the input data. Obvious manipulations are intentionally mislabelling the datasets, labeling a picture of a cat as a dog for instance, or reducing, possibly increasing, the number of images of a category in a dataset. The mechanics of the data curation process ought to facilitate the discovery of such manipulations. The author is unaware of any publicly available datasets secured with even primitive methods.

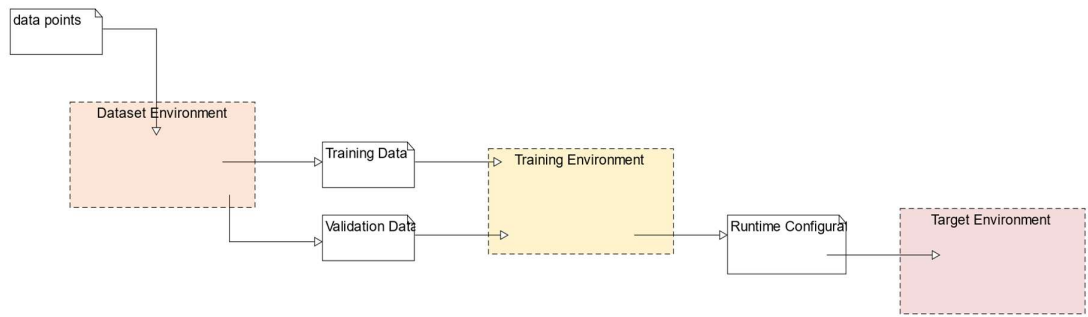

Fig. 1: Environments and their Interfaces in a typical Neural Network Workflow

---

[1] It is to be noted that the paper itself underwent some corrections [9] and the doi appears to be no longer valid [10]

Goldblum et. al provide a structured articulation of the issues concerning dataset integrity [14].

More problematic is the perturbation problem. It is possible to introduce perturbations into a multi-dimensional data-input such as an image, which are not necessarily visible to the human eye and hence not detectable. These perturbed images can be used to drastically manipulate the learning process; the results of a classification run [4] or, as it turns out, discover characteristics of the classifier [15], in other words steal the model. These three attack scenarios clearly illustrate not only robustness issues with the machine learning technology model [16]–[18] but also represent vulnerabilities in a security sense [19] and integrity issues in a safety sense [20].

*2) Modelling Environment:* The dataset environment prepares data for the modelling (training) and validation phase. The modelling environment can be viewed as a configurable tool. This comparison allows us to view the modelling environment to a certain extent as a variant of the model-based design domain and refer to well-known solutions and methods of dealing with vulnerabilities in this domain.

The operation of this tool – in the naïve case – poses some issues. In an example from the pynq environment from Xilinx [21] for the MNIST number-recognition, the tooling begins with random extraction of validation and training data from the data set. It then, for every training epoch, randomises the order in which the training data is presented. These two randomisations result in a model that is practically irreproducible. Whilst the recognition rates are similar, what is worrying from a safety point of view is that the set of false positives is different. This can be seen in figure where we show the results of passing the entire MNIST data set through a trained network (see Fig. 2). Whilst the representation is in logarithmic scale to enhance the display of false positives (in red) it can be seen that the profiles differ – we train a network on two similar platforms and let the training go through 1000 epochs – and the number and position (relative to true positives) are different. Determinism in training is not, in the naïve form, given.

The only ML-specific attack on the training process not associated with manipulation of the input data that the author is aware of, utilises involve perturbations on the tool. These include gradient perturbation and objective function perturbation [22]. Ironically these perturbations can also be legitimately applied to the tooling to make the model more robust against probing-based discovery.

From a safety point of view the integrity of the modelling environment is critical. Under the assumption that the correct architecture has been chosen and the data sets reflect what they are supposed to, the tooling will release some file that can be used to initialise and perform the MAC calculations on a specific processing architecture (CPU or GPU or the like). In any functional safety environment such modelling software will require qualification to ensure the output is deterministic and traceable. One such tool, TensorFlow Lite [23], generates a runtime that requires Linux as an operating system. In this case the tool is not qualified; it may be possible to qualify the runtime; but Linux is not qualified. Another tool, Keras [24], generates json files describing the architecture and a weights file which [25] used to generate C-code which can then be analysed using static code analysis. Such a tool has the potential to be qualified, but Keras itself would also require qualification to be used in a functional-safety environment.

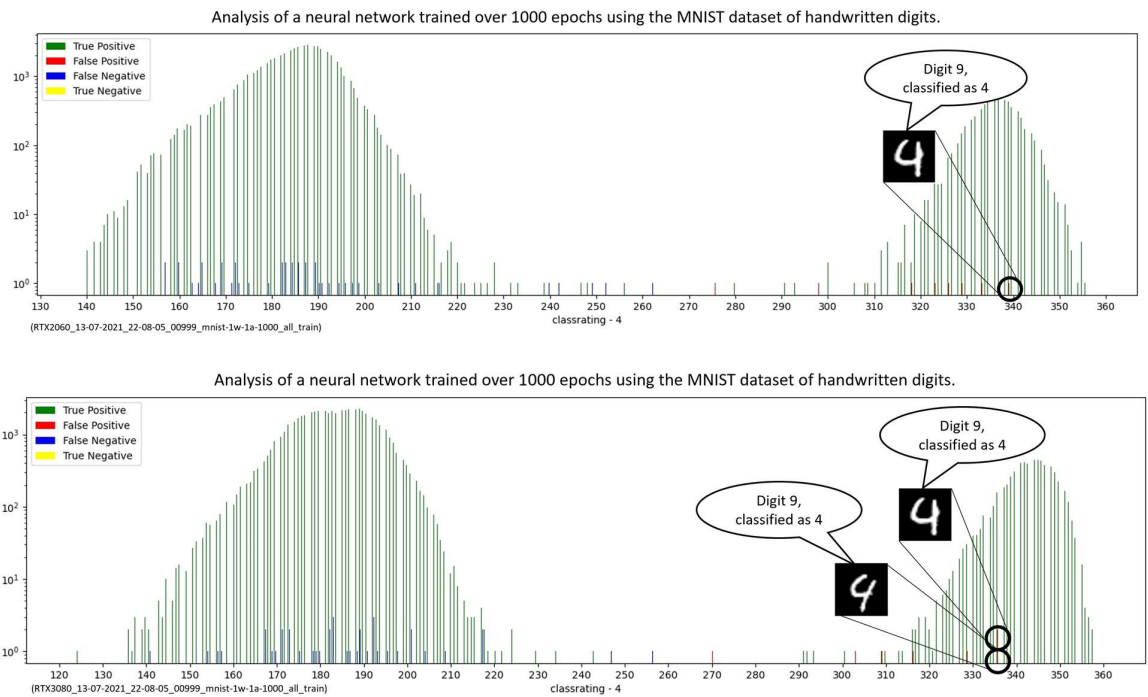

Fig. 2: Two density distributions generated by applying all elements of the MNIST data base to binary models trained on the data set on a logarithmic scale. The training ceases after 1000 epochs/iterations. The distributions clearly show different numbers and different relative positions of false positives. Note: whether the common false positive is in truth a four or a nine s purely in the eyes of the data-labeller.

*3) Target Environment:* By target environment is meant the hard- and software platform upon which the MAC calculations are performed. Massively parallel computing architectures, such as GPUs or application-specific accelerators, are the architectures of choice. The typical unit of execution which runs on a GPU is a kernel, a sequence of operations which can be conceptually encapsulated as a thread. The kernel is passed to the GPU with an indication of the number of required threads and their organisation – 32 threads per block and 16 blocks would correspond to 512 iterations of a for-loop on a CPU architecture. These blocks are then scheduled on a compute unit of the GPU by a scheduler resident on the GPU and from there to processing elements by a scheduler resident in the compute unit. By and large the execution of a neural network classifier model on hardware is just another program/process and as such, at least in terms of safety, allows us to refer back to well-understood analytic and implementation mechanisms for ensuring the execution proceeds as it is ought.

*a) Safety:* The well-known method for ensuring that program execution is correct – that is it performs as the source code would suggest – is a (tightly coupled) lockstep architecture [26] in which the code is executed on two processors simultaneously and the execution – code fetch, data fetch and data writeback – is compared on a system-bus level. Should the two processors execution differ this is taken as an unrecoverable error and a reset is asserted. Loosely coupled architectures also exist where the results of a calculation are compared rather than its execution.

There is an expectation that the execution time, both worst case and best case, should be known as timing integrity is considered important. Memory integrity is also considered important.

*b) Security:* There are attack modes that focus on model theft namely model extraction attacks and model inversion attacks. Model extraction attacks aim to steal the intellectual property of the model by querying a model (viewed as black-box) with inputs designed to model the model as a set of equations and hence duplicate the parameters and with it the (original) model [27]. Model inversion attacks are designed to expose the privacy of records used in the training of the model, for instance medical records. A complex theme, [22] provides a good starting point for further perusal.

*c) Execution Schedule Integrity:* One could reasonably expect, in a world where constant-time multipliers are the norm, a relatively time-constant execution behaviour. Any naïve experiment – for instance as shown in the kernel execution-time measurements of a simple vector addition (Fig. 3 clearly demonstrates a statistical distribution) – will convince one that this is not necessarily the case. GPU kernel execution performance – usually a single threading multiple data execution model (STMD) - is a function of thread structure, memory location of variables (register, local, shared or global memory) tasks and the organisation and mapping of threads onto the hardware. Using OpenCL terminology, a kernel is composed of a number of threads (e.g. iterations of a loop) organised into blocks of threads which are executed on a processing element. The blocks are assigned to a compute unit which organises a number of processing elements and forms part of a compute device. As the compute device (GPU) lives in an asynchronous relationship with an attached host (CPU), and the number of threads/blocks generally is greater than the number of processing elements, a GPU-based scheduler is required to schedule execution of the blocks on the compute elements and the threads on the processing elements. The scheduler introduces behaviour that is not, in the face of incomplete documentation, transparent. The measurements of Fig 3 are arrived at through the use of Nvidia's nvprof host-based profiler so inaccuracies are to be expected. In the context of this discussion, this means the implementer has to go further to ensure his code executes with time-integrity.

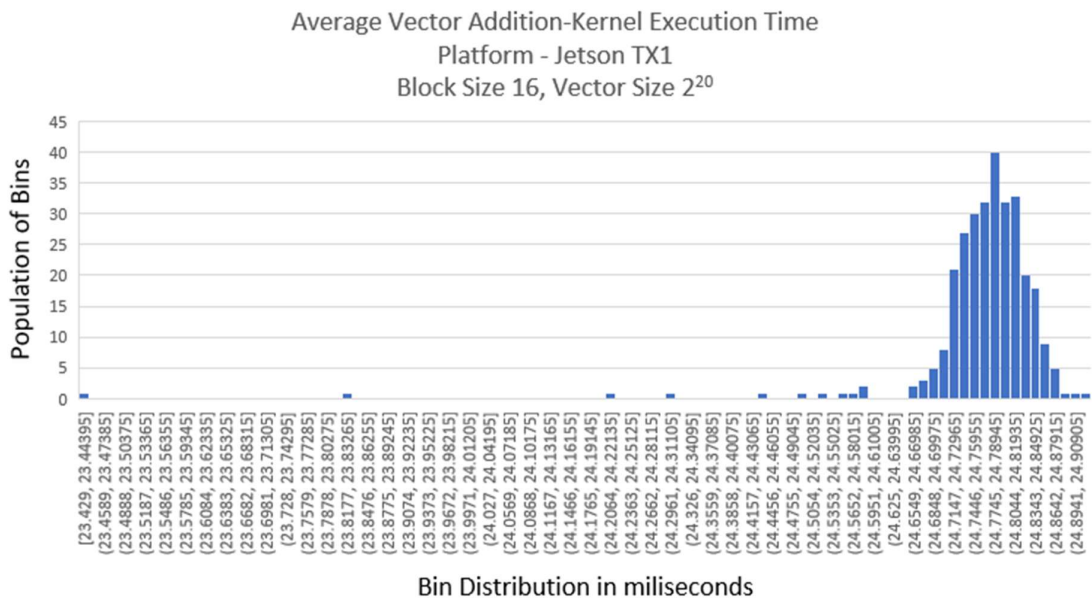

Fig. 3: The execution time density distribution of a simple GPU, vector-addition, kernel as measured by Nvidea's nvprof tool executed 300 times. The kernel was written in Python and the execution time includes the host-sided activity of the JIT compiler. Regardless of the technology used, we wish to illustrate the nature of a statistical distribution, in particular, in this case, the extreme best-case execution time.

This unclarity has not gone unnoticed by others [28]. The major chip manufacturers usually offer, under non-disclosure agreements, access to GPU interfaces that can be used to provide runtime safety guarantees. The only other way the programmer has to influence the order of execution is by scheduling the host-sided issuance of kernels. Concurrent execution of different kernels is considered a challenge [29] and in fact this has been used to construct covert-channel attacks [30].

*d) Memory Integrity:* Memory integrity is fundamental in terms of both security and safety. In particular the use of dynamically allocated memory is something of an issue in both safety and security contexts and neural network representations are known to require significant memory. This is especially a concern in IoT-class embedded devices where memory-scheduling has been proposed to alleviate strictures imposed by a constrained platform [31].

The clearly delineated and architecture-agnostic memory model of OpenCL systems (see Fig. 4) makes a good reference point to show the potential complexity of memory issues in terms of safety and security. Unsurprising there are well documented attack modes such as buffer-overflow and information-leakage attacks possible on all levels of memory [32].

*e) Platform Integrity:* In terms of platform integrity we must consider whether the model is being executed properly, that is whether the calculations have been performed correctly. As previously mentioned, tightly-coupled lockstepping architectures are common in industry and well-understood. Tightly coupled lockstepping requires duplication of expensive computing resources, especially so in the context of GPUs. We therefore expect that loosely-coupled lockstepping will become the norm in this technology-domain and some work has been done in this area [33] but results are as of yet inconclusive.

III. CONCLUSION

Embedded-AI, more specifically embedded neural network-based classifiers are in the process of moving from a buzzword to a common-place reality. These systems represent part a typical embedded system and part a novelty embedded system in that we have architectures that are in general common to many advanced embedded circuits but differ in particular in some aspects. In addition, the workflow required to arrive at an embedded-AI solution is substantially different to a typical code-based embedded system but bears strong similarities to model-based design. We can therefore intuit that in many cases we can use best practices well-understood from known architectures and configurations. Our contribution consists of identifying the exceptions where special care and indeed some applied-research/research needs to be carried out.

One area where the embedded-AI technology differs from model-based design is the endless manipulation possibilities of the dataset. This begins with data bias which can be manipulated to ensure the model classifies incorrectly. Then there is the wide-reaching perturbation issue, one whose effects are neither fully understood nor completely researched. Slight perturbation of input data has been used to corrupt models, it has been used to produce wildly faulty classifications and has been used to steal models. As of yet, the author is not aware of any silver-bullet solution, but has observed, from the digital image authentication domain, that many verification tools use other signal-domains, frequency, for instance as inputs rather than amplitude as is common [34], there seems to be little reason why these two domains could not be run in parallel with the second domain acting as a sort of cage for results from the first. This observation is also supported by additional and related observations from the digital image watermarking domain where the camera supplying images can be identified both by model and by product instance (serial number). Otherwise, it could appear that the workflow in applying security mitigations applicable to model-based design systems could equally be applied to the process of making a typical embedded-AI design flow.

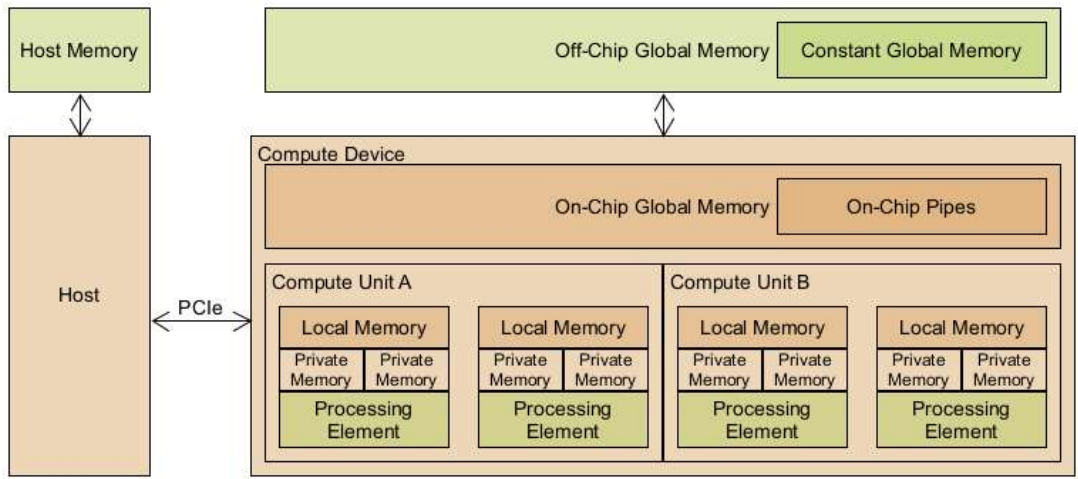

Fig. 4: The generic OpenCL model as applied to a Compute Device. The model clearly shows the precise categorisation of memory types as well as the asynchronous communication between host and compute element (PCIe).

Achieving a safety certification is a more challenging task as this requires demonstration of functional correctness and demonstration of safety-relevant quality requirements and neither can currently be guaranteed. The very point of using machine learning classifiers is because implementers are unable to otherwise define the features that should effect a classification. In cases like these, the safety certification authorities are put in the difficult position of not wanting to be seen to stand in the way of legitimate innovation and so support implementers in finding an acceptable caging strategy for mitigating misclassifications. On the other hand, there is neither scope nor reason for turning a blind eye to obvious process and execution defects. Once the determinism issue is solved we can expect that tool qualifications will follow.

As it stands, we do not fully understand neural network classifiers, we cannot guarantee robustness of a neural networks, and we can certainly not guarantee deterministic behaviour. There is however sufficient research work going on in this domain for us to expect a better understanding and so move in this direction. While the big chip developers have begun to integrate architectural features in their execution engines (CPUs, GPUs and other accelerating architectures), there is little to no openly available literature about these features so work on generic solutions is warranted.

ACKNOWLEDGMENTS

Thanks are due to Thorvin Stasiak for creating and preparing the trained networks and their predictions in Figure 2, rendering Figure 4, and formatting the paper and to Gabriel Wicki for his review. Any errors are those of the author.